\documentstyle[11pt,epsf]{l-aa}
\def\arcmin              {$^{\prime}$}

\def\arcsec              {$^{\prime\prime}$}

\def\kmsmpc              {km\thinspace s$^{-1}$\thinspace Mpc$^{-1}$}

\begin{document}
\thesaurus{}
\title{The geometry  of the    quadruply imaged  quasar   PG~1115+080;
implications   for $H_0$\thanks{Based  on  observations taken   at the
Nordic Optical Telescope and the Canada France Hawaii Telescope}}
\author{F. Courbin\inst{1,2}, P. Magain\inst{1\thanks{Also Ma\^{\i}tre
de     Recherches    au    FNRS  (Belgium)}},   C.R.   Keeton\inst{3},
C.S.  Kochanek\inst{3},  C. Vanderriest\inst{2}, A.O.  Jaunsen\inst{4}
and J. Hjorth\inst{5}}
\thesaurus{12.07.1; 12.03.3; 11.17.4; 13.19.3}
\offprints{F. Courbin (Li\`ege address)}
\institute{
Institut d'Astrophysique, Universit\'e de Li\`ege,
Avenue de Cointe 5, B--4000 Li\`ege, Belgium
\and
URA 173 CNRS-DAEC, Observatoire de Paris,
F--92195 Meudon Principal C\'edex, France
\and
Harvard-Smithsonian Center for Astrophysics, MS-51, 60 Garden Street,
Cambridge, MA 02138, USA
\and
Institute of Theoretical Astrophysics, University of Oslo,
Pb. 1029, Blindern, 0315 Oslo, Norway
\and
NORDITA, Blegdamsvej 17, DK--2100 Copenhagen {\O}, Denmark\\}
\date{Submitted ; Accepted}
\maketitle
\markboth{F. Courbin et al.: Geometry of PG~1115+080}{}
\begin{abstract}
\looseness=-8 Time delay measurements have  recently been reported for
the lensed   quasar PG~1115+080.  These  measurements   can be used to
derive $H_0$, but only if we can constrain  the lensing potential.  We
have applied a recently  developed deconvolution technique to  analyse
sub-arcsecond $I$ band images  of PG~1115+080, obtained at the  Nordic
Optical Telescope (NOT) and the Canada France Hawaii Telescope (CFHT).
The  high performance of the   deconvolution code allows us to  derive
precise positions  and magnitudes  for the four   lensed images of the
quasar, as well as for the lensing galaxy.  The new measurement of the
galaxy position  improves  its precision by a  factor  of $3$ and thus
strengthens the  constraints on the lensing  potential.   With the new
data, a range of models incorporating some of the plausible systematic
uncertainties yields $H_0 = 53^{+10}_{-7}$ \kmsmpc.


\end{abstract}

\section{Introduction}

Gravitationally lensed quasars  with multiple images provide a  unique
tool for determining  the cosmological parameters independently of the
classical methods.  In particular, time delays between the images of a
lensed quasar   can  be used to  determine  the  Hubble constant $H_0$
(e.g. Refsdal 1964, Grogin \& Narayan 1996).

Extensive photometric monitoring of lensed quasars has been undertaken
to obtain  accurate   light curves for  the most    promising objects.
Unfortunately,  precise time delays have  been difficult to obtain due
to poor temporal sampling  of the light curves, possible contamination
by microlensing and short time scale events. 

\looseness=-4 Light  curves   have  recently  been obtained  for   the
quadruply imaged  quasar PG~1115+080  (Schechter  et al.   1997).  The
good time coverage, together with the  fact that microlensing does not
seem to significantly contaminate the intrinsic photometric variations
of the  lensed source, yield  measurements of two  time delays between
three   components  of the system  (Schechter   et al.  1997, Bar-Kana
1997).  Keeton and Kochanek (1997, hereafter KK97)  showed that it was
difficult   to  constrain $H_0$ because  of  degeneracies  in the lens
models.   However, KK97 also  showed   that the constraints  could  be
improved by  measuring the position  of the lensing galaxy relative to
the quasar images better than was  possible with the pre-refurbishment
HST observations  of Kristian et al.   (1993, hereafter K93).  The aim
of the  present paper  is to  provide  a better  determination of  the
galaxy position and, thus, an improved estimate of $H_0$.

\section{Observations}

The observations consist of five $I$ band images of PG 1115+080 with a
typical seeing of 0\farcs6-0\farcs7.  Using SIS, the tip-tilt adaptive
optics camera installed at the Cassegrain focus  of the CFHT, two 600s
exposures were  taken on the night of  1995 December 27.  The detector
was  a Loral 3,  2048$\times$2048 pixel CCD (pixel scale 0\farcs0865).
In addition, three $I$ band frames were  obtained at the NOT in direct
imaging mode.  Two 200s exposures were taken on the night of 1996 June
7 and an equivalent   exposure was obtained  on  1996  June  10.   The
Brocam2    2048$\times$2048  CCD   camera   was   used   (pixel  scale
0\farcs1071).

\section{Deconvolution - a new algorithm}

\begin{figure}[t]
\begin{center}
\leavevmode 
\epsfxsize= 8.0cm 
\epsffile{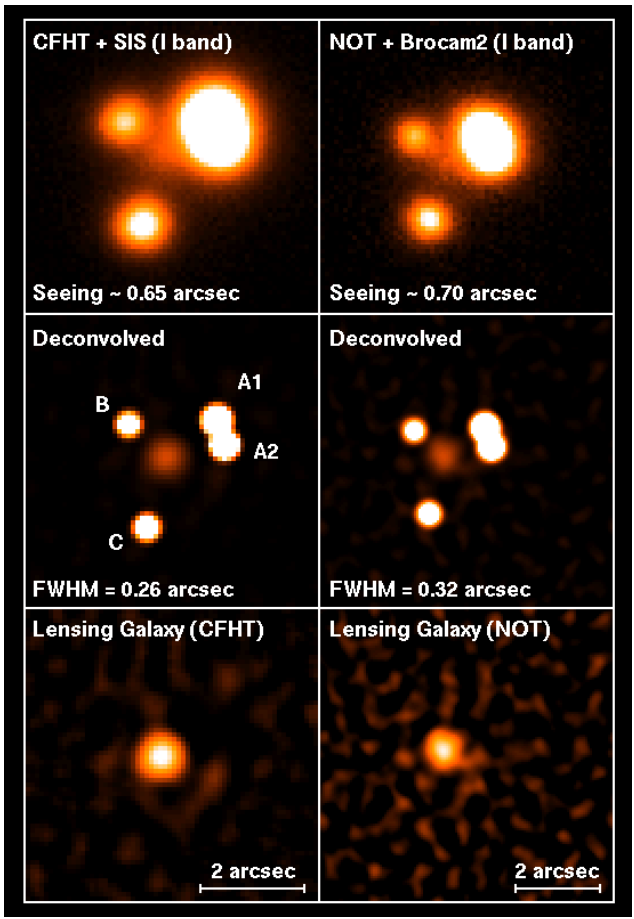}
\vspace*{4mm}
\caption{Top panels: I band images of PG~1115+080 obtained at the CFHT
(600s  exposure, field  of 5\farcs5) and   at the NOT  (200s exposure,
field of 6\farcs8).  Middle  panels: the same images,  but deconvolved
using our new algorithm. Bottom panels: the deconvolved lensing galaxy
alone. Note that the greyscale was  chosen to display the full dynamic
range of the images.}
\end{center}
\end{figure}

\looseness=-6  The  images   were  deconvolved using   a  new  method,
described by  Magain,   Courbin \& Sohy  (1997),   that allows precise
photometric and   astrometric measurements  of strongly  blended point
sources superimposed on a diffuse  background.  The main idea of  this
algorithm is to  deconvolve by a PSF narrower  than the total observed
PSF so as to preserve a good sampling of the  deconvolved image.  As a
consequence, 1) the image can be decomposed in a sum of point sources,
plus diffuse background and 2) the final resolution of the deconvolved
images is {\it chosen} by the user.

%The main idea is that the deconvolution should not use the total Point
%Spread Function  (PSF),    because this   tries to   recover   Fourier
%frequencies  higher   than the Nyquist    frequency and  violates  the
%sampling  theorem.  The resulting ``deconvolution artefacts'' preclude
%any quantitative measurements on the  deconvolved frame.  Instead, one
%should use a  narrower PSF, which yields   images of improved but  not
%infinite  resolution that  are   compatible with  the {\it  adopted\/}
%sampling size of the deconvolved image.
%\looseness=-6 
%In other words, the  deconvolved image has its   own PSF which can  be
%{\it chosen}  by  the user.  The  shape of  any  point source  in  the
%deconvolved image  is thus known  {\it a priori}  and  this {\it prior
%knowledge} can be used to decompose the deconvolved model image into a
%sum of point sources plus a diffuse  background.  The best model image
%is  computed by minimizing  the $\chi^{2}$ using  an algorithm derived
%from the conjugate gradient  method.  The positions and intensities of
%the point sources  and the image of  the deconvolved background galaxy
%are direct byproducts of the deconvolution.

\looseness=-6 We chose   the final deconvolved PSF of   the CFHT as  a
Gaussian  with a Full-Width-Half-Maximum  (FWHM) of 3 pixels.  The NOT
images were deconvolved on a grid of pixels two times smaller than the
original frames (i.e.  0\farcs05355) in  order to keep a good sampling
of the lensing galaxy, and the final PSF was a Gaussian with a FWHM of
6 (smaller) pixels.  This led to resolutions of 0\farcs26 for the CFHT
and  0\farcs32 for the NOT.   The PSF used  to reach these resolutions
was constructed  from two stars  about one magnitude brighter than the
quasar image and situated 1\arcmin\ away from the lens.

\looseness=-6 The weight  attributed  to the  local smoothing of   the
background  component  in our deconvolved model   image (see Magain et
al.\ (1997) for more details) was chosen so that  the residuals at the
location    of the   lensing  galaxy   had   the correct   statistical
distribution, i.e.\ Gaussian with a standard deviation  of 1 (in units
of the  photon noise).  In this  way we avoided under- or over-fitting
in the area  of interest. A constant smoothing  term was used  for the
whole image and was adapted  to the area  of the lensing galaxy.  This
leads to  slight overfitting of  the  sky noise  farther away from the
target and produces some noise amplification at low light levels.

\looseness=-6 In the   CFHT images, the PSF  shows  variations of  the
order of   5--10\% from one  side  of the total   field to  the other.
Therefore, we allowed  the PSF to  depart  slightly from  that derived
from the two neighboring stars.  The PSF was constrained to deviate by
the smallest amount consistent  with an artefact-free deconvolution of
the 4 quasar images.

\looseness=-6 Fig.~1 shows the result of the  deconvolution for a CFHT
and a NOT image.  Since  the greyscale was  chosen to display the full
dynamic  range,   this   image  clearly   shows  that  there   are  no
``deconvolution artefacts'' or  ``ringing  effects'' around  the point
sources.  

\section{Geometry of PG~1115+080}

Five images were used to measure the position of the lens galaxy.  For
each  frame, the deconvolution procedure  returned  the coordinates of
the four point sources, the deconvolved image  as well as the image of
the  diffuse objects  (Fig.~1).  The latter  was  used to  measure the
position of the lensing galaxy.

\looseness=-4 To   compare   our   results  with K93's     results, we
transformed our coordinates  to the coordinate  system they used.  The
scaling factor  and rotation angle were  chosen to match the positions
of the four point sources as well  as possible, and the transformation
was then applied to the galaxy coordinates.  Fig.   2 compares the new
positions with  the positions from  K93.  The precision  of the galaxy
position has  been increased by more  than a factor  of 3, from 50 mas
with the  pre-refurbishment HST images (K93), to  15 mas  with the new
data and deconvolution technique.  Note that K93 used a pixel scale of
0\farcs04389, which  Gould \&  Yanny  (1994) revised  to 0\farcs04374.
Table\  1\  summarizes   the  geometry of   PG~1115+080,   in the same
orientation as K93,  but using the pixel  size given by Gould \& Yanny
(1994).

\begin{figure}[t]
\begin{center}
\leavevmode
\epsfysize=7.0 cm
\epsffile{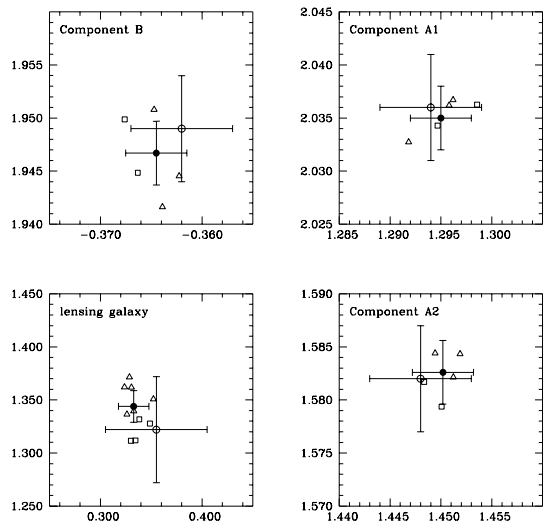}
\caption{Positions  of components A1,   A2, B and the lensing  galaxy,
relative to component C.  The scale for  the point sources is 10 times
larger than for the galaxy.  Triangles and squares indicate individual
measurements   from the  NOT and CFHT    images, respectively.  Filled
circles with error bars show our  final results from the weighted mean
of the individual measurements.  Open circles with error bars show the
HST positions from K93.}
\end{center}
\end{figure}

The galaxy position was measured on each deconvolved background by the
first two authors independently, using either  Gaussian fitting or the
first  order moments   of  the light  distribution, resulting   in  10
measurements (one  for each author and  each frame). We first averaged
the two measurements for each frame and  then computed a weighted mean
of the 5  independent measurements, with a  weight twice as large  for
the NOT images as for the CFHT images because the latter are corrupted
by PSF variation across the field.

The   standard deviation of the   mean is $\sim2$   mas  for the point
sources and  $\sim10$  mas for the lensing   galaxy.  The effect  of a
variable PSF was  tested by   performing numerical simulations   which
indicate that  possible   systematic effects  could affect  the  point
sources by another 2  mas and the  lensing galaxy by an  additional 10
mas.  This leads to final 1$\sigma$ error bars of  3 mas for the point
sources and 15 mas for the lensing galaxy.

%Numerical  simulations  show    that  no
%systematic error larger than  10  mas should affect  our measurements.
%The most obvious error would come from a variation of the wings of the
%PSF between the A1 and A2 quasar images and the stars used for the PSF
%determination.  However, a check on the few stars present in the field
%does not indicate the presence of such an effect.

The intensity ratio of A2  relative to A1  is given in  Table 1.  This
ratio does not show significant variations among our frames or between
our frames and the K93  frames, even though  they  were taken at  four
different    epochs.    This suggests    that  microlensing  does  not
significantly affect the light curves of A1 and A2 in the $I$ band.

The  galaxy  magnitude was  derived  on the NOT  deconvolved frames by
aperture photometry (0\farcs9 diaphragm  in diameter).  The zero point
was computed  using several standard stars.  Images with a much higher
signal-to-noise would be necessary to  derive the shape of the galaxy.
However, the galaxy observed in the present data is compatible with a
fuzzy circular object, broader than a point source.

\begin{table}[t]
\begin{center}
\caption{Relative  positions of the  four lensed images of PG~1115+080
relative to component C (in arcseconds). The intensity ratios in the I
band are given for the NOT observations obtained on  the night of 1996
June 7. The magnitude of A1 is $I=16.34\pm0.05$}
\begin{tabular}{c c c c}
\hline \hline
Image & $I/I_{A1}$ & $X$ (\arcsec)      &     $Y$ (\arcsec)  \\
\hline  
A1  & $1.000 \pm 0.000$  &   $+1.291 \pm  0.003$  &  $+2.028 \pm 0.003$\\
A2  & $0.683 \pm 0.010$  &   $+1.445 \pm  0.003$  &  $+1.578 \pm 0.003$\\
B   & $0.164 \pm 0.003$  &   $-0.364 \pm  0.003$  &  $+1.940 \pm 0.003$\\
C   & $0.255 \pm 0.005$  &   $+0.000 \pm  0.000$  &  $+0.000 \pm 0.000$\\
Lens& $I = 19.6 \pm 0.1$ &   $+0.332 \pm  0.015$  &  $+1.339 \pm 0.015$\\
\hline
\end{tabular}
\end{center}
\end{table}

\section{Implications for $H_0$}

Schechter et al.  (1997) recently  obtained light curves for images B,
C, and   the mean of  the close  pair A$=$A1+A2  (Fig.~1).  Bar-Kana
(1997) analysed the light curves using  a variety of assumptions about
how to treat the photometric errors.  He found  that the best resolved
time delay is $\Delta\tau_{BC}=25.0_{-1.7}^{+1.5}$  days, and that the
other independent time  delay is best  expressed in terms  of the time
delay       ratio       $r_{ABC}\equiv\Delta\tau_{AC}/\Delta\tau_{BA}=
1.13_{-0.17}^{+0.18}$, with a $\sim0.2$ systematic uncertainty.

Using the  quasar image and galaxy  data  of K93, Keeton,  Kochanek \&
Seljak  (1997),  Schechter et al.\   (1997), and  KK97  found that the
system can be fit only by including  a perturbation from a small group
of galaxies near the lens galaxy (Young  et al.  1981, Kundi\'c et al.
1997).  There are significant degeneracies   in the models related  to
the position of the group and to the profiles of the galaxy and group,
but with more precise data the degeneracies can be reduced (KK97).  In
particular, reducing the uncertainties in  the lensing galaxy position
can constrain the position of the group.

\looseness=-4 We  studied the effects of  the improved  measurement of
the galaxy position  by recomputing the  models from KK97.  We modeled
the   galaxy as an  ellipsoidal   mass  distribution with a   variable
position, ellipticity, and orientation, and we used both an isothermal
(dark matter) model  and a constant  $M/L$ model.   The constant $M/L$
model was  approximated by a modified Hubble  model because  it has an
analytic  deflection  formula  and is  therefore   simpler  than  a de
Vaucouleurs $r^{1/4}$ model;  the  Hubble model  was chosen  to have a
fixed  core radius $s=0\farcs2$ ($0.55   h^{-1}$ kpc).  We modeled the
group  as a spherical mass  distribution with a variable position, and
we  considered both a singular isothermal  sphere (SIS) model expected
for a dark matter halo, and a point mass  model to examine the effects
of making the group  more concentrated. Increasing  the core radius of
the galaxy or group would decrease the  implied value for $H_{0}$, but
dark matter distributions generally appear to be singular.
 
\begin{figure}[t]
\begin{center}
\leavevmode
\epsfysize=7.0 cm
\epsffile{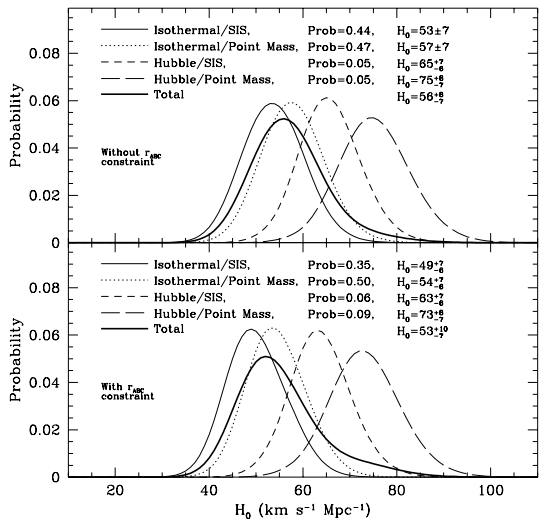}
\caption{\looseness=-4 Normalized  probability distributions for $H_0$
computed  from a  Bayesian analysis   with the  four  classes of  lens
models, and a  total probability distribution  taking into account the
relative  probabilities.  In the top  panel the Bayesian analysis does
not include the constraint from the time delay ratio $r_{ABC}$, and in
the bottom panel it does use $r_{ABC}$.  The relative probabilities of
the four classes of  models, and their implied  values for  $H_0$, are
given in the key.  This Figure is to be compared with Fig.~6 of KK97.}
\end{center}
\end{figure}

Table~2 compares the new  results with the  results from KK97 to  show
the   effects of improving the  galaxy  position.  For  models with an
isothermal galaxy the best-fit $\chi^2$ has decreased despite the fact
that  the  constraints have become  stronger, while  for models with a
Hubble galaxy the $\chi^2$ has increased slightly but the fit is still
good.

\looseness=-4 Following KK97,   we  determined $H_0$ from   a Bayesian
analysis of the lens models.  Using discrete  model classes limits the
generality  of the  analysis, but until   even better constraints  are
available (such as the  shape and profile of the  galaxy) a full model
survey is unwarranted. The four model classes examined here illustrate
the range of effects from model uncertainties; for other model classes
see   KK97.   By reducing  the range   of  acceptable lens models, the
improved constraint from the galaxy position decreases the $H_0$ error
bars by a factor of two (see Table 2 and  Figure 3).  The new position
also increases the probability of  the isothermal (dark matter) galaxy
relative    to  the  constant  $M/L$  galaxy.     Finally, it slightly
strengthens the discrepancy noted by KK97 between  the lens models and
Bar-Kana's  (1997) value   for $r_{ABC}$.    Improved measurements  of
$r_{ABC}$  would reveal whether  the  apparent  discrepancy is due  to
systematic effects in the measurements or to a poor lens model.
 
\begin{table}[t]
\begin{center}
\caption{Model  results,  including   the   absolute $\chi^2$ of   the
best-fit model  (with $N_{dof}=3$) and  the values for $H_0$  from the
Bayesian analysis including the constraint  from the time delay  ratio
$r_{ABC}$.  Quantities in parentheses  are results from KK97 using the
data from K93.}
\begin{tabular}{r c c c}
\hline\hline
           & Galaxy:   &   Isothermal    &  Modified Hubble \\
\hline
SIS Group  & $\chi^2$  & $1.35$ ($1.77$) &  $2.09$ ($1.47$) \\
           &  $H_0$    & $49_{-6}^{+7}$ ($44\pm11$) & $63_{-6}^{+7}$ ($55\pm14$) \\
Point Mass & $\chi^2$  & $1.32$ ($1.83$) &  $1.95$ ($1.70$) \\ 
Group      &  $H_0$    & $54_{-6}^{+7}$ ($51\pm11$) & $73_{-7}^{+8}$ ($64_{-17}^{+16}$) \\ 
\hline
\end{tabular}
\end{center}
\end{table}

%Figure~3 and Table~2 show the normalized probability distributions and
%summarize the implied values for $H_0$.  For each class of models, the
%improved constraint from the galaxy position has significantly reduced
%the range of acceptable values for $H_0$, decreasing the error bars by
%roughly a factor of  2.  In addition, the new  data have increased the
%relative   probability of the  isothermal (dark  matter) model for the
%galaxy.  With the K93 data a dark matter galaxy was more likely than a
%constant $M/L$ galaxy  by a little more than  $2:1$, but  with the new
%data the  ratio  is  almost $6:1$.  Finally,    the new data  slightly
%strengthen the  discrepancy noted by KK97  between the lens models and
%Bar-Kana's (1997)  value  for $r_{ABC}$.    Improved measurements   of
%$r_{ABC}$   should reveal whether  the  apparent discrepancy is due to
%systematic effects in the measurements;  if not, then the measurements
%of $r_{ABC}$ might  rule out current lens models  and provide a strong
%constraint on new models.

%Finally, KK97 noted  that a constant  $M/L$ galaxy is more  consistent
%than an isothermal  galaxy with Bar-Kana's  (1997) value for  the time
%delay ratio $r_{ABC}$.  Since the new quasar  and galaxy data strongly
%favor a dark matter  galaxy, there seems  to be a  growing discrepancy
%between the  models and  Bar-Kana's  (1997) value for   $r_{ABC}$.  At
%present the  discrepancy is not  serious because Bar-Kana's result has
%significant systematic uncertainties,   but  with improved   precision
%$r_{ABC}$ may provide a strong constraint on lens models.

Based on K93 galaxy position, KK97 found $H_0=51_{-13}^{+14}$ \kmsmpc.
The  Bayesian analysis  using our  new data  gives a total probability
distribution with  $H_0=53_{-7}^{+10}$ \kmsmpc, where these error bars
incorporate  the time delay  uncertainties  as well as some systematic
uncertainties in the lens models related to the profiles of the galaxy
and group and to the position of the group.

Our new deconvolution  technique  could be used   to analyse all   the
images  used for the light curves  and to derive the independent light
curves of A1 and A2.    The high precision   of the method might,   in
addition, make it possible to further  narrow the uncertainties on the
time delays and hence on $H_0$.

\begin{acknowledgements}
%      ~~~~~~~~~~~~~~~~ 
F.C. is supported  by ARC 94/99-178  ``Action de Recherche Concert\'ee
de la  Communaut\'e Fran\c{c}aise'' and P\^ole d'Attraction
Interuniversitaire  P4/05  (SSTC, Belgium).  C.R.K.   is  supported by
ONR-NDSEG grant N00014-93-I-0774.   C.S.K. is  supported by NSF  grant
AST-9401722.
\end{acknowledgements}

\begin{thebibliography}{}

\bibitem{} Bar-Kana R., 1997, preprint, astro-ph/9701068

\bibitem{} Gould A., Yanny B., 1994, PASP 106, 101

\bibitem{} Grogin N.A., Narayan R., 1996, ApJ 464, 92; erratum
1996, ApJ 473, 570

\bibitem{} Keeton C.R., Kochanek C.S., 1997, in press 

\bibitem{} Keeton C.R., Kochanek C.S., Seljak U., 1997, ApJ 482, 604

\bibitem{} Kristian J. et al. 1993, AJ 106, 1330

\bibitem{} Kundi\'c T., Cohen J.G., Blandford R.D., 1997, 
preprint, astro-ph/9704109

\bibitem{} Magain P., Courbin F., Sohy S., 1997, preprint, astro-ph/9704059

\bibitem{} Refsdal S., 1964, MNRAS 128, 295  

%\bibitem{} Refsdal S., 1964b, MNRAS 128, 307

\bibitem{} Schechter P.L. et al. 1997, ApJ 475, L85

\bibitem{} Young, P.J., Deverill, R.S., Gunn, J.E., Westphal, J.A., 
Kristian, J., 1981, ApJ, 244, 723

\end{thebibliography}
\end{document}